\documentclass{JAC2003}

\usepackage{graphicx}
\usepackage{booktabs}
\usepackage{array}
\usepackage{subfigure}

\begin{document}
\title{STUDY OF BEAM DIAGNOSTICS WITH TRAPPED MODES IN \\THIRD HARMONIC SUPERCONDUCTING CAVITIES AT FLASH\thanks{Work supported in part by the European Commission under the FP7 Research Infrastructures grant agreement No.227579.}\\[-.8\baselineskip]}

\author{P.~Zhang$^{1,2,3,}$\thanks{pei.zhang@desy.de}, N.~Baboi$^3$, R.M.~Jones$^{1,2}$, I.R.R.~Shinton$^{1,2}$\\
\mbox{$^1$School of Physics and Astronomy, The University of Manchester, Manchester, U.K.}\\
\mbox{$^2$The Cockcroft Institute of Accelerator Science and Technology, Daresbury, U.K.}\\
\mbox{$^3$Deutsches Elektronen-Synchrotron, DESY, Hamburg, Germany}}

\maketitle

\begin{abstract}
Of{}f-axis beams passing through an accelerating cavity excite dipole modes among other higher order modes (HOMs). These modes have linear dependence on the transverse beam of{}fset from the cavity axis. Therefore they can be used to monitor the beam position within the cavity. The f{}ifth dipole passband of the third harmonic superconducting cavities at FLASH has modes trapped within each cavity and do not propagate through the adjacent beam pipes, while most other cavity modes do. This could enable the beam position measurement in individual cavities. This paper investigates the possibility to use the f{}ifth dipole band for beam alignment in the third harmonic cavity module. Simulations and measurements both with and without beam-excitations are presented. Various analysis methods are used and compared. A good correlation of HOM signals to the beam position is observed.
\end{abstract}

\section{Introduction}
An electron bunch excites wakef{}ields when passing through an accelerating cavity. These wakef{}ields can be decomposed into higher order modes (HOM), whose transverse components are usually dominated by dipole modes~\cite{rwake}. Dipole modes have a linear relation to the transverse beam position of the excitation bunch. Therefore, the beam of{}fset within the cavity can be determined by monitoring the dipole modes. The principle has been proved on 1.3~GHz TESLA cavities \cite{racc1-3}, and we intend to apply the idea to the third harmonic cavities \cite{rhom-1}.

\begin{figure}[h]
\centering
\includegraphics[width=0.47\textwidth]{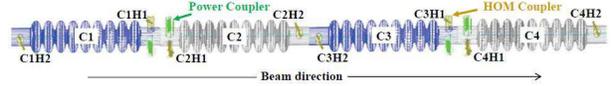}
\caption{Schematic of the four cavities within ACC39.}
\label{cavity-cartoon}
\end{figure}

Third harmonic cavities operating at 3.9~GHz were installed in FLASH \cite{rflash} at DESY to linearize the energy spread of the electron bunch induced by the main accelerating module containing TESLA cavities for bunch compression \cite{racc39-1}. There are four inter-connected 3.9~GHz cavities in the cryo-module ACC39 (Fig.~\ref{cavity-cartoon}). Due to the small size of the 3.9~GHz cavity itself (scaled down from 1.3~GHz cavity by a factor of 3) and the larger-diameter beam pipes connecting the cavities (larger than one-third of those of the 1.3~GHz cavity), the HOM spectrum is much more complex than that of the 1.3~GHz cavity \cite{rhom-3,rhom-4}. Previous studies have shown that the dipole beampipe modes and the f{}irst two dipole passbands respond well to the beam movement \cite{rhom-6}. This paper focuses on the f{}ifth dipole passband.

\section{Trapped Modes in the Fifth Dipole Band}
Most of the dipole cavity modes couple to adjacent cavities through attached beam pipes. However, there are modes in the f{}ifth dipole band trapped within each cavity. A simulation is performed on an ideal cavity without couplers using CST Microwave Studio \cite{rcst}, with a convergence of 0.0001\%, 1.3~million mesh cells and electric boundary conditions. The resulting f{}ield distributions are shown in Table.~\ref{simu-table} along with the mode frequencies and R/Qs. The f{}irst 4 modes seems trapped but weak couplings to the beam are expected because of small R/Q values.
\begin{table}[h]
\setlength\tabcolsep{4pt}
\centering
\caption{The Fifth Dipole Band from the Simulation}
\medskip
\begin{tabular}{ccc}
\toprule
\textbf{E-f{}ield} & \textbf{f}(GHz) & \textbf{R/Q}($\Omega$/cm$^2$)\\
\midrule
\includegraphics[width=0.25\textwidth]{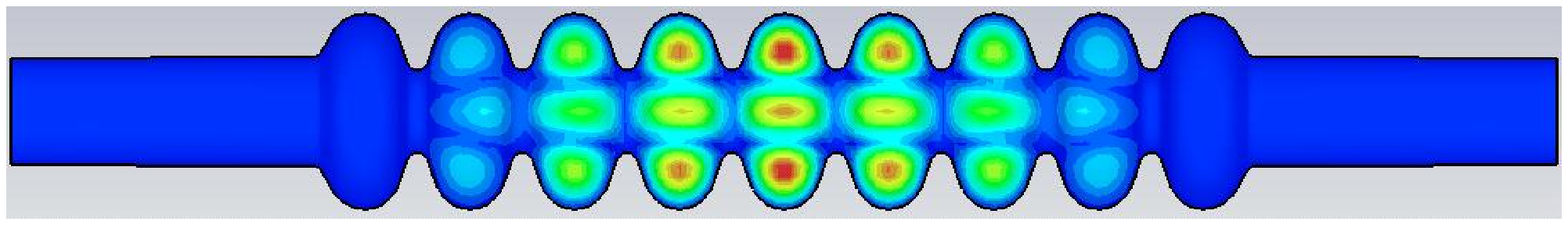}    & 9.052    & 0.00\\
\midrule
\includegraphics[width=0.25\textwidth]{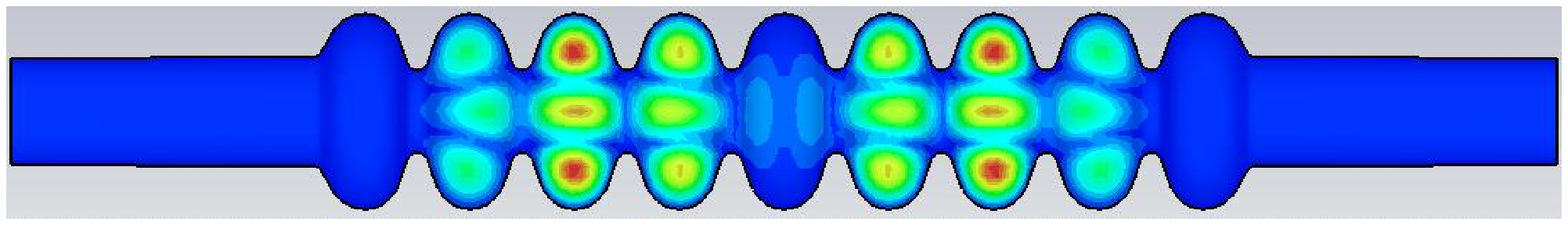}    & 9.053    & 0.05\\
\midrule
\includegraphics[width=0.25\textwidth]{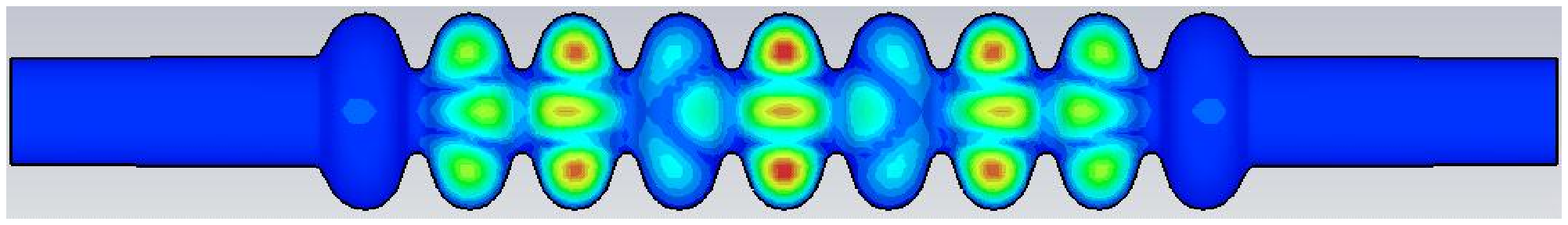}    & 9.055    & 0.06\\
\midrule
\includegraphics[width=0.25\textwidth]{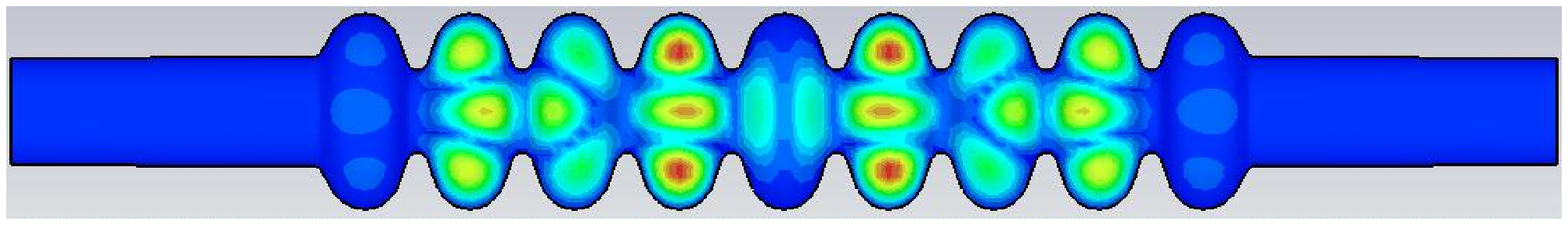}    & 9.058    & 2.17\\
\midrule
\includegraphics[width=0.25\textwidth]{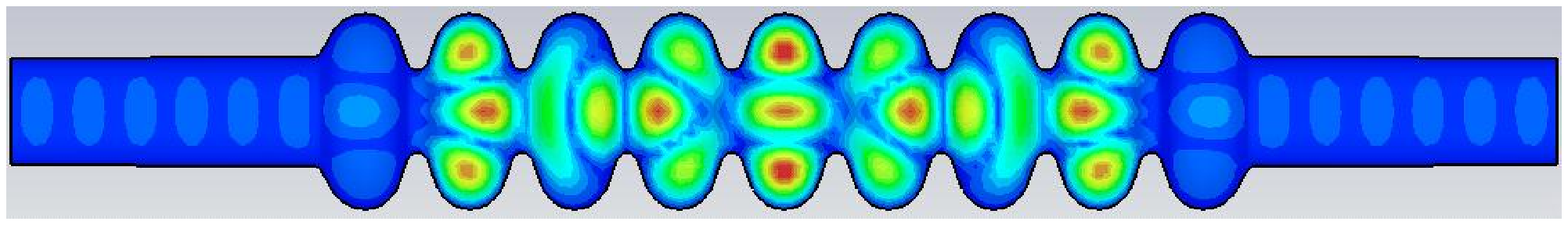}    & 9.066    & 4.12\\
\midrule
\includegraphics[width=0.25\textwidth]{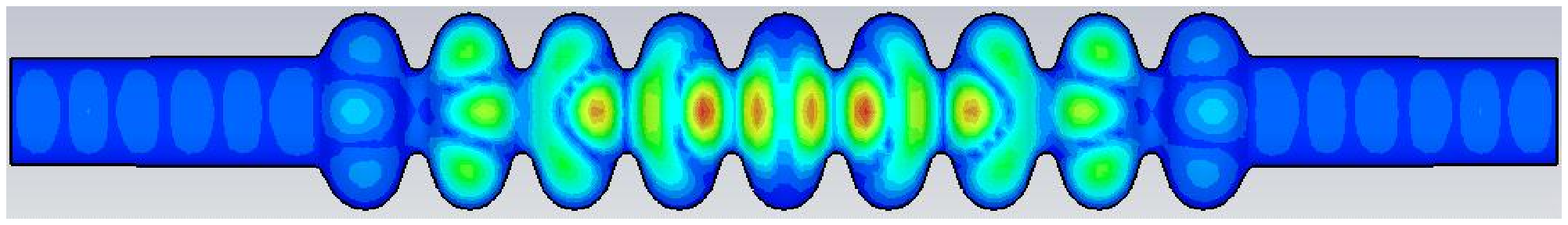}    & 9.089    & 0.58\\
\bottomrule
\end{tabular}
\label{simu-table}
\end{table}

\begin{figure}[h]
\centering
\includegraphics[width=0.46\textwidth]{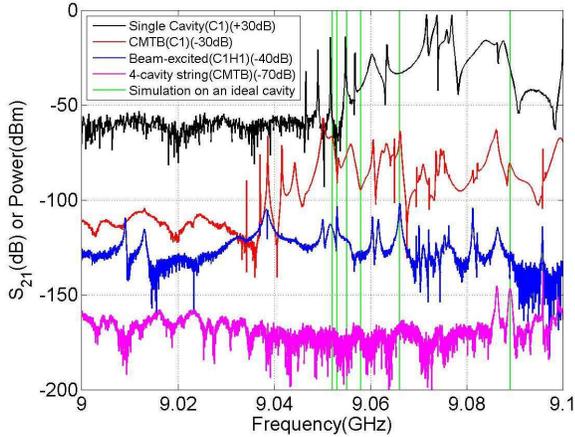}
\caption{The f{}ifth dipole passband.}
\label{rsa-cmtb-fnal-C1}
\end{figure}

To examine whether the modes are trapped in real cavities, various measurements were conducted both with and without beam-excitation. The transmission spectra measured at the single cavity test stand in Fermilab have been analyzed\footnote{Data are provided by T.~Khabibouline from Fermilab.}. The result for C1 is shown in Fig.~\ref{rsa-cmtb-fnal-C1} (black curve). The mode frequencies deviate from simulations because the cavity is not ideal and the couplers break the symmetry of the ideal structure. Also, the cavity was not tuned to the accelerating mode frequency (3.9~GHz). Then the spectrum measured across C1, after assembly in the cryo-module at CMTB (Cryo-Module Test Bench)\footnotemark[\value{footnote}], is shown as red curve in Fig.~\ref{rsa-cmtb-fnal-C1}. The modes shift in frequency and/or change in shape, due to cavity tuning, etc. To test the mode localization, the spectra are compared with the measurement across the entire four-cavity string (from C1H2 to C4H2, magenta curve in Fig.~\ref{rsa-cmtb-fnal-C1}). Comparing with the single cavity spectrum (red curve in Fig.~\ref{rsa-cmtb-fnal-C1}), modes below 9.08~GHz do not propagate. These modes are of great interest for our study, because they enable the beam position to be determined within each cavity. Beam-excited spectra were measured in FLASH by using a Real-time Spectrum Analyzer (RSA) from all eight HOM couplers (blue curve in Fig.~\ref{rsa-cmtb-fnal-C1}). The spectrum from 9 to 9.05~GHz was excited by a single electron bunch with a charge of about 0.5~nC, while the one from 9.05 to 9.1~GHz was excited by another single electron bunch with almost the same beam properties. Compared with CMTB spectra (red and magenta curves in Fig.~\ref{rsa-cmtb-fnal-C1}), modes excited by the beam are in general consistent with those from transmission measurements, while some modes were not picked up by the HOM coupler or simply not excited by the beam.

Both simulations and measurements show that some modes in the f{}ifth dipole band are trapped below 9.08~GHz. As the spectrum above 9.05~GHz is excited by a single bunch, the spectrum from 9.05 to 9.08~GHz is used for the following analysis.

\section{Response of Dipole Modes to Beam Movement}
\subsection{Measurement Scheme}
\begin{figure}[h]
\includegraphics[width=0.48\textwidth]{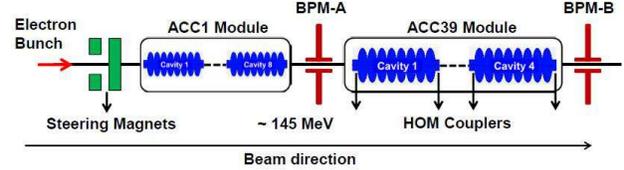}
\caption{Schematic of measurement setup (not to scale, cavities in ACC1 are 3 times larger than those in ACC39).}
\label{hom-setup}
\end{figure}

To investigate the response of dipole modes to the transverse beam position, we moved the beam to create various transverse of{}fsets in the cavity. The measurement setup is shown in Fig.~\ref{hom-setup}. Two steering magnets were used to kick the beam horizontally and vertically. The bunch (about 0.5~nC) was subsequently accelerated by ACC1, then went through ACC39 with various transverse of{}fsets. The position within each cavity is obtained by interpolating the readouts from two beam position monitors (BPM-A and BPM-B in Fig.~\ref{hom-setup}), which are situated on each side of ACC39. A straight-line trajectory of the beam was attained by switching of{}f the accelerating f{}ield in ACC39 and quadrupole magnets nearby.

\subsection{Dipole Dependence by Lorentzian F{}it}
\begin{figure}[h]
\subfigure[Mode amplitudes (C2H2)]{
\includegraphics[width=0.23\textwidth]{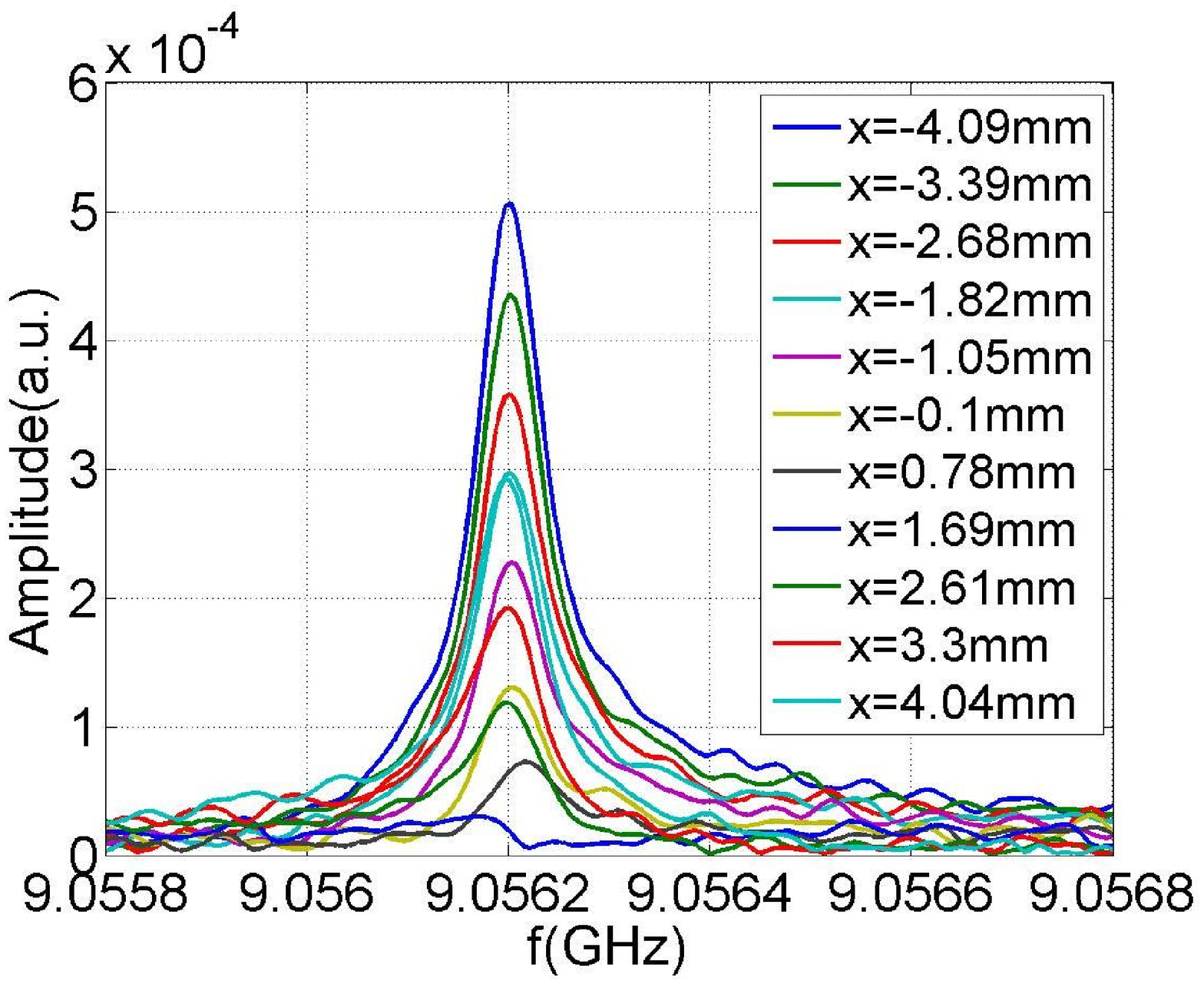}
\label{amp-cross-D5}
}
\subfigure[Linear dependence]{
\includegraphics[width=0.23\textwidth]{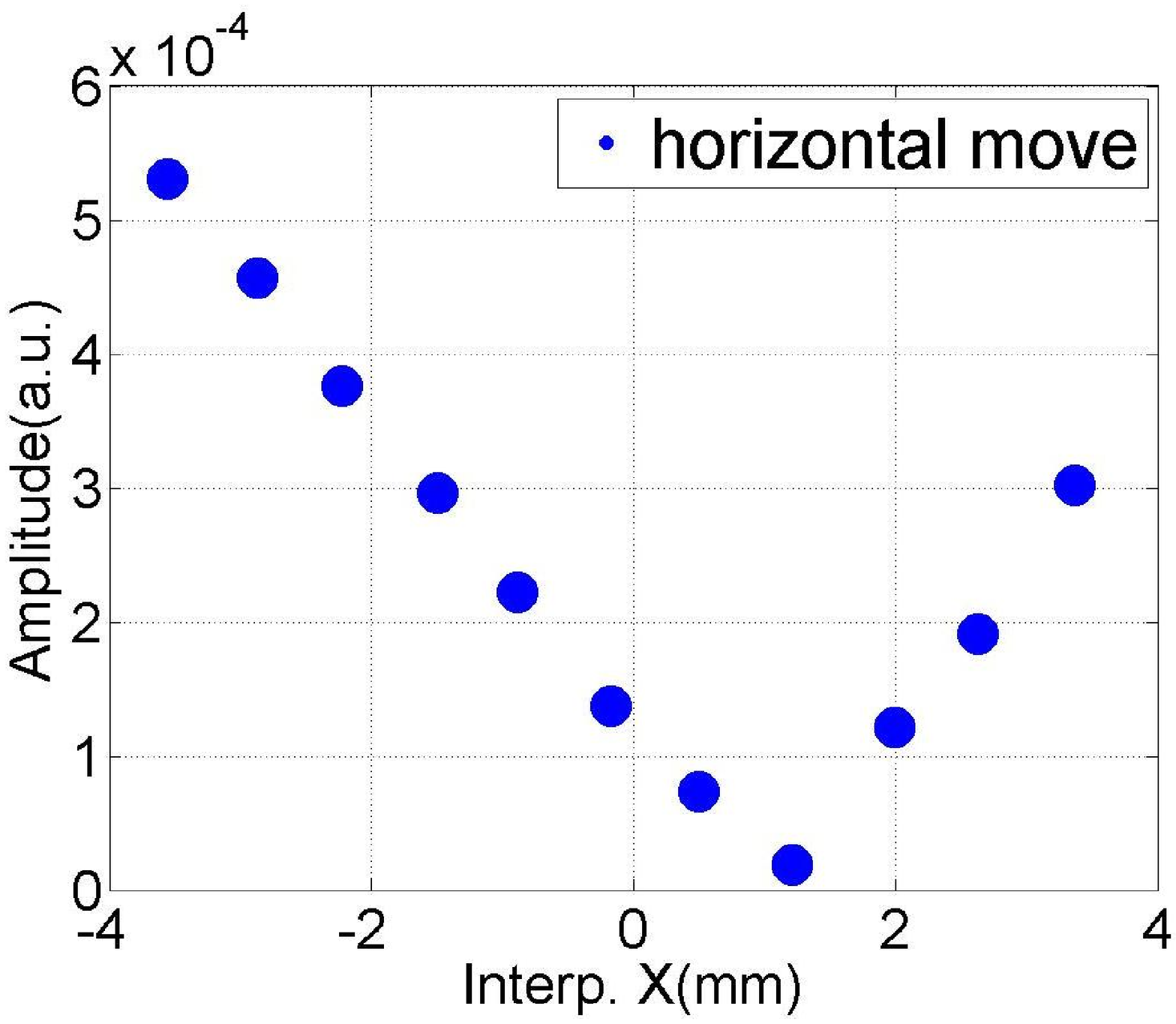}
\label{dep-cross-x-D5}
}
\caption{Linear dependence of mode amplitude on the transverse beam position.}
\label{amp-dep-cross-D5}
\end{figure}

Initially, the beam was kicked horizontally without changing the vertical steerer. The mode amplitude varies with the horizontal of{}fset, as shown in Fig.~\ref{amp-cross-D5} for one mode at about 9.0562~GHz measured from C2H2. The vertical position in C2 varies by $\pm$0.29~mm. By exploiting a Lorentzian f{}it, the amplitude of each peak in Fig.~\ref{amp-cross-D5} is obtained and plotted against the horizontal beam position in Fig.~\ref{dep-cross-x-D5}. Linear dependence of the mode amplitude on the transverse beam position can be observed, which is a dipole-like behavior.

\begin{figure}[h]
\subfigure[Grid move]{
\includegraphics[width=0.23\textwidth]{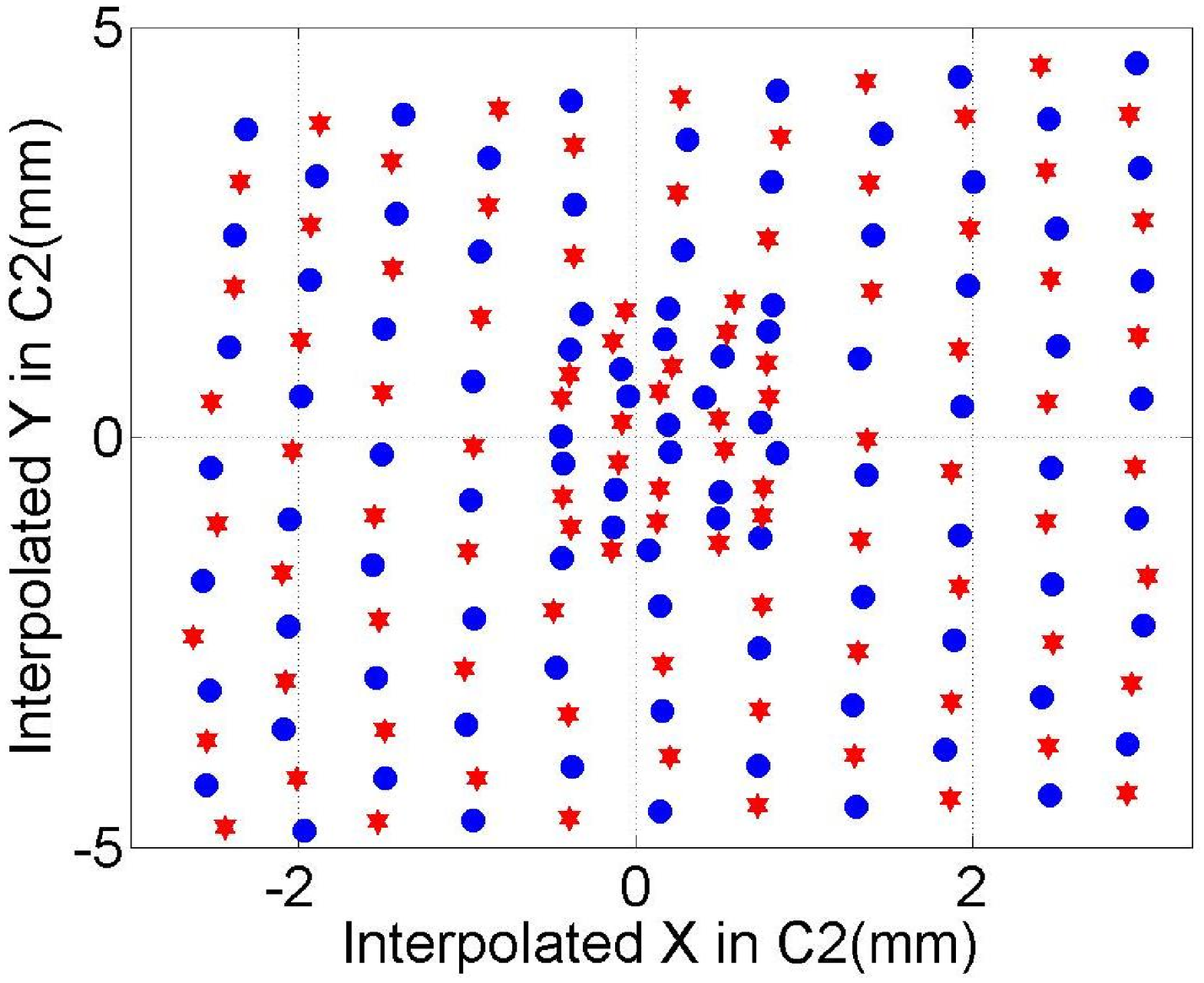}
\label{2D-grid}
}
\subfigure[Polarization (C2H2)]{
\includegraphics[width=0.23\textwidth]{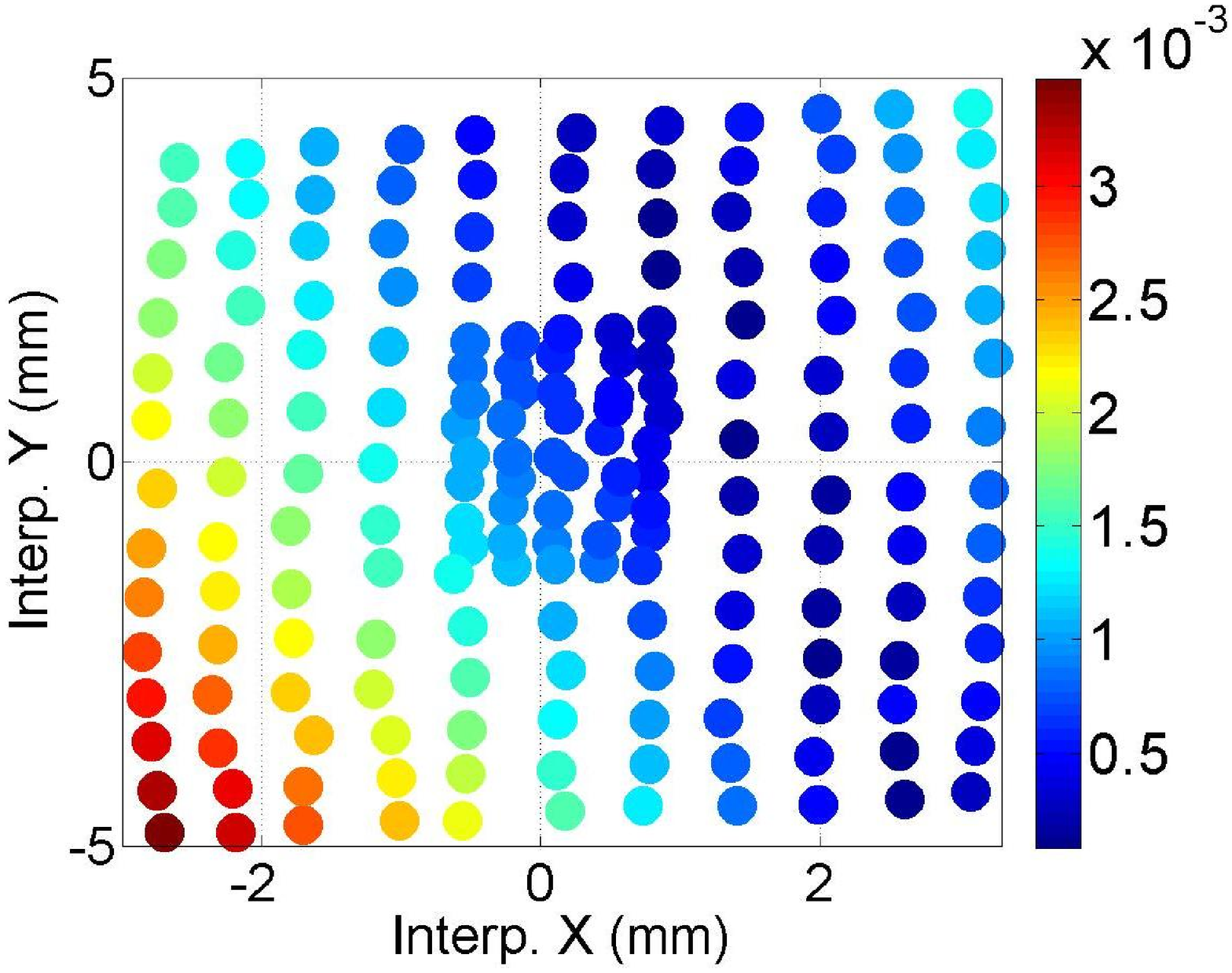}
\label{dep-grid-D5}
}
\caption{Beam movement and mode polarization.}
\label{2D-grid-dep-D5}
\end{figure}

The beam was then moved in a grid-like manner, as shown in Fig.~\ref{2D-grid}. The tilt is attributed to coupling between the $x$ and $y$ planes caused by the ACC1 module and the two BPMs. The data are split into calibration samples (blue dots) and validation samples (red asterisks) for further analysis. The amplitude of the same mode is again obtained by Lorentzian f{}it for each beam position, and plotted in Fig.~\ref{dep-grid-D5}. The color denotes the amplitude magnitude. The polarization of this mode can be observed.

\subsection{Dipole Dependence by DLR and SVD}
\begin{figure}[h]
\subfigure[Spectrum (C1H1)]{
\includegraphics[width=0.23\textwidth]{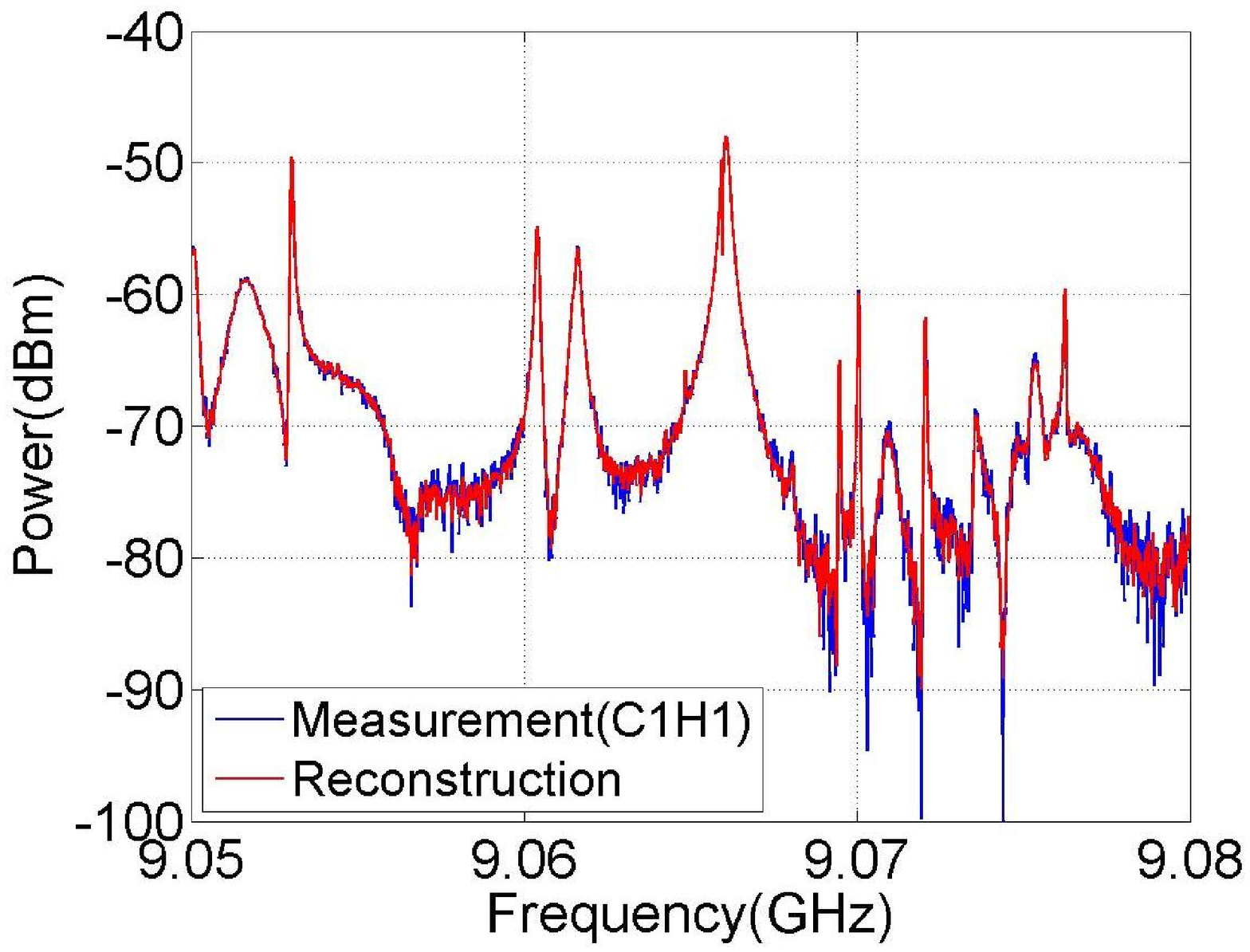}
\label{svd-reco}
}
\subfigure[Singular values]{
\includegraphics[width=0.23\textwidth]{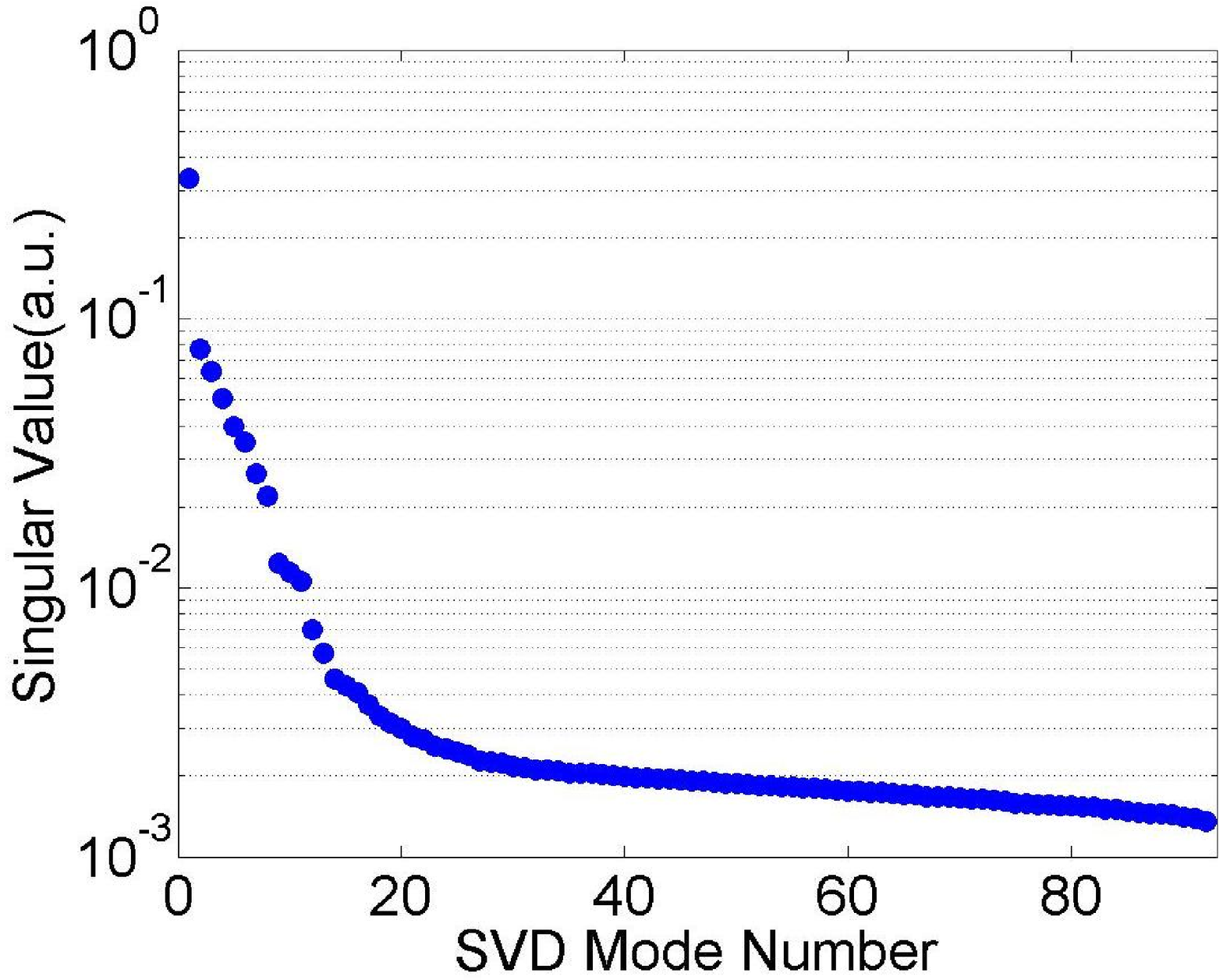}
\label{svd-sv}
}
\caption{Spectrum measured from C1H1 and singular values decomposed from calibration samples.}
\label{svd-reco-sv}
\end{figure}

To study the dipole linear dependence, a \textit{direct linear regression} (DLR) is applied on the spectra (Fig.~\ref{svd-reco}) by
\begin{equation}
A\cdot M + B_0 = B,
\label{eq-dlr}
\end{equation}
where each row of matrix $A$ denotes a spectrum amplitude in linear scale (normalized by beam charge) taken at one beam position, each row of matrix $B$ denotes one set of transverse beam position in $x$ and $y$, and $B_0$ is an of{}fset matrix. The linear system is solved by least-squares method. The calibration samples (see Fig.~\ref{2D-grid}) are used to build the matrix $M$ and then tested on validation samples.

\begin{figure}[h]
\subfigure[Calibration samples]{
\includegraphics[width=0.48\textwidth]{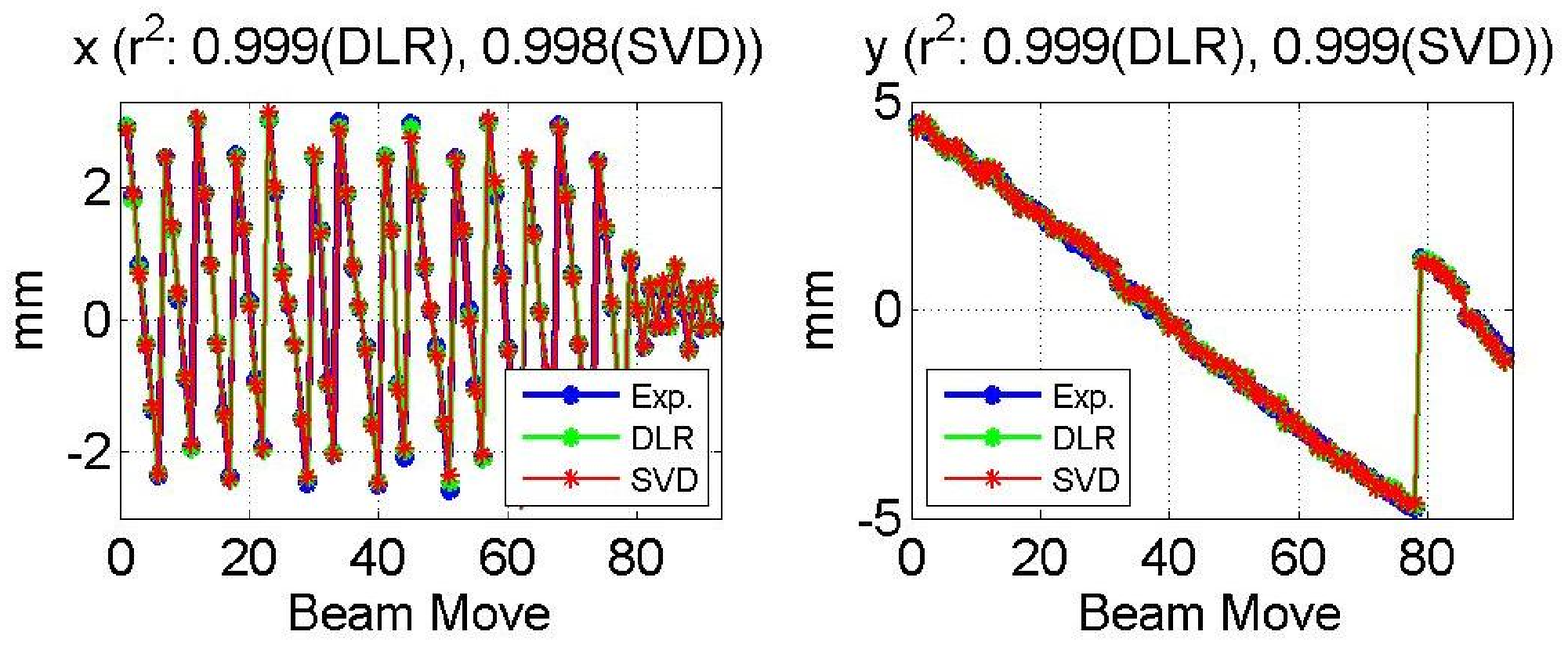}
\label{Bo-Bop}
}
\subfigure[Validation samples]{
\includegraphics[width=0.48\textwidth]{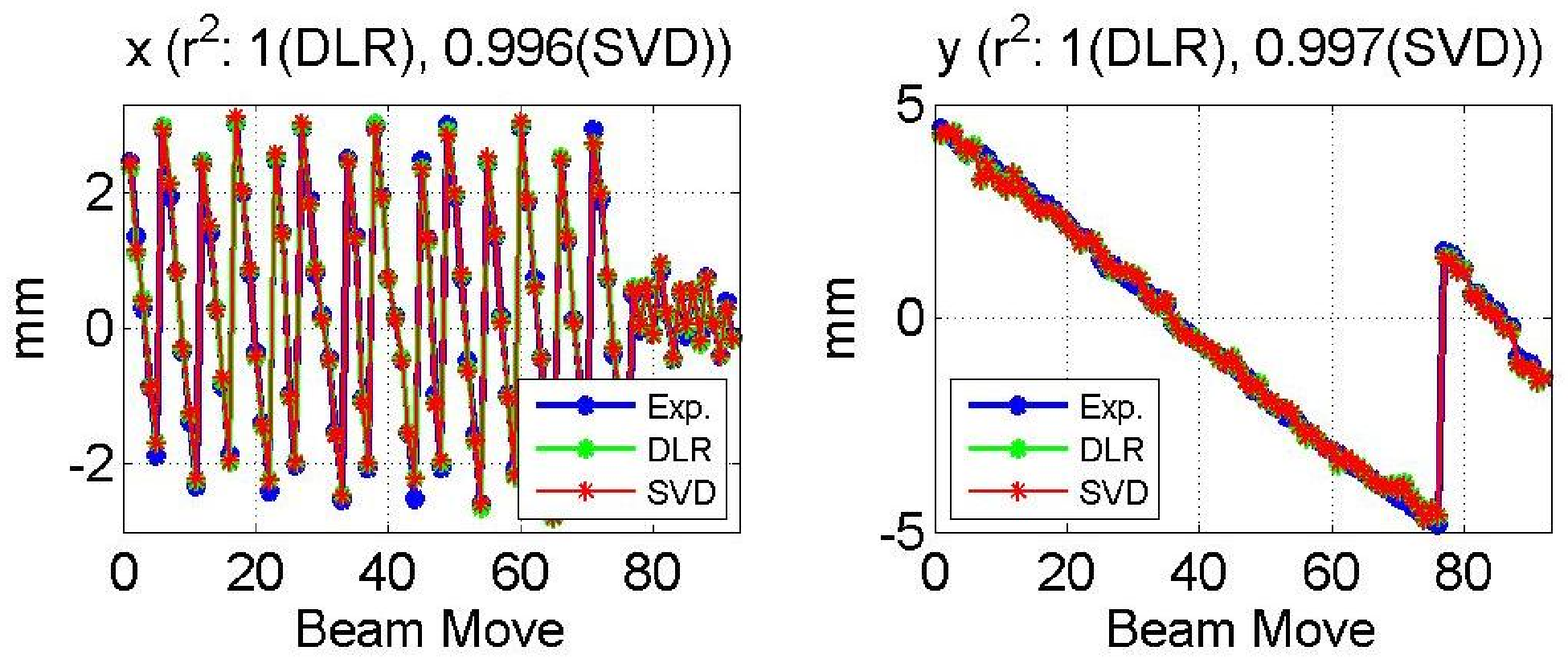}
\label{Bv-Bvp}
}
\caption{Transverse beam position predicted by DLR (green) and SVD (red, using the f{}irst twenty SVD modes). Measurements are in blue. $r^2$ is def{}ined as in \cite{rstat-1}.}
\label{B-Bp}
\end{figure}

The number of sampling points determines the size of matrix $A$ in Eq.~\ref{eq-dlr}. A huge number of samples rapidly makes the calculation of $M$ computationally very expensive. Therefore, the \textit{singular value decomposition} (SVD) \cite{racc1-3} method is alternatively applied to reduce the system size by looking for the prominent patterns of matrix $A$ in terms of SVD modes:
\begin{equation}
A=U\cdot S\cdot V^T,
\label{eq-svd}
\end{equation}
where $U$ and $V$ are formed by base vectors, and $S$ contains the singular values (Fig.~\ref{svd-sv}). Using only the f{}irst twenty SVD modes according to their singular values, one can recover the spectra used for decomposition (Fig.~\ref{svd-reco}). Then linear regression is applied on the SVD amplitude matrix formed by the f{}irst twenty SVD modes instead of the much larger sized matrix $A$. The results are shown in Fig.~\ref{B-Bp} for both the DLR and the SVD method. Excellent agreement is attained with measurements for both methods. This provides conf{}idence that the system can be well represented by a few SVD modes.

\newpage

\section{Conclusions}
Some modes in the f{}ifth dipole band of the third harmonic cavities have been shown to be localized within each cavity by both simulations and measurements. According to spectra characteristics, various techniques are applied and compared: Lorentzian Fit, DLR and SVD. Linear dependence of the mode amplitude on the transverse beam position is observed, comf{}irming the dipole character of the modes. These modes are considered in the development of HOM electronics for beam diagnostics, along with propagating cavity modes in the f{}irst two dipole passbands and localized beampipe modes.


\end{document}